\documentclass[12pt]{article}
\usepackage{amsmath,amssymb,latexsym}
\usepackage[dvips]{graphicx}
\providecommand{\beqa}{\begin{eqnarray}}

\providecommand{\tit}{\textit}
\providecommand{\mrm}{\mathrm}
\providecommand{\eeqa}{\end{eqnarray}}
\providecommand{\ti}{{\mathrm{t}}}
\def\Orb{{\mathbf{S}^1/\mathbf{Z}_2}}
\def\Z2{{\mathbf{Z}_2}}
\def\mX{{\mathbf{X}}}

\begin{document}

\thispagestyle{empty}
\rightline{HUTP-06/A0025}

\addtolength{\baselineskip}{1.5mm}

\thispagestyle{empty}

\vspace{32pt}

\begin{center}

\textbf{\Large Power Spectrum and Signatures
for Cascade Inflation}\\

\vspace{32pt}

Amjad Ashoorioon$^{a,}$\footnote{amjad@astro.uwaterloo.ca} and
Axel Krause$^{b,c,}$\footnote{akrause@fas.harvard.edu}

\vspace{32pt}

\textit{${}^{a}$Department of Physics, University of Waterloo,
Waterloo, Ontario, N2L 3G1, Canada}\\
\textit{${}^{b}$Jefferson Physical Laboratory, Harvard University,
Cambridge, MA 02138, USA\\
${}^{c}$George P. \& Cynthia W. Mitchell Institute for Fundamental
Physics, \\
Texas A\&M University, College Station, TX 77843, USA}

\end{center}

\vspace{32pt}

\begin{abstract}

\addtolength{\baselineskip}{1.2mm}

The power spectrum of M-theory cascade inflation is derived. It
possesses three distinctive signatures: a decisive power suppression at small scales, oscillations around the scales that cross the horizon when the inflaton potential jumps and stepwise decrease in the scalar spectral index. All three properties result from features in the inflaton potential. Cascade inflation realizes assisted inflation in heterotic M-theory and is driven by non-perturbative interactions of $N$ M5-branes. The features in the inflaton potential are generated whenever two M5-branes collide with the boundaries. The derived small-scale power suppression serves as a possible explanation for the dearth of observed dwarf galaxies in the Milky Way halo. The oscillations, furthermore, allow to directly probe M-theory by measurements of the spectral index and to distinguish cascade inflation observationally from other string inflation models.

\end{abstract}

\setcounter{footnote}{0}

\newpage

\section{Introduction}

For a long time it seemed difficult to connect inflation to
string-theory. In its low-energy approximation string-theory is
described by supergravity. Inflation based on the F-term potentials
of 4-dimensional ${\cal N}=1$ supergravities resulting from
string/M-compactifications suffers from a large slow-roll parameter $\eta$. The origin of this problem traces back to the appearance of the K\"ahler-factor $\exp(K)$ in the F-term potential. New possibilities to address this problem arose with the advent of
D-branes \cite{Polchinski:1995mt}. They allowed to identify the inflaton with open string modes such as the geometrical distance between two D-branes \cite{Dvali:1998pa}.

An inflaton requires a very shallow potential. Hence, a priori,
moduli serve as natural candidates. To provide them with a non-trivial potential supersymmetry needs to be broken which can be done in various ways. One might add anti D-branes to the open string sector \cite{Kachru:2003sx} or supersymmetry breaking fluxes to the closed string sector \cite{Dine:1985rz}. Also the inclusion of non-perturbative instanton effects leads to spontaneous supersymmetry breaking in the low-energy supergravity \cite{Curio:2001qi}. Assuming just a single inflaton, the task for deriving inflation from string-theory becomes then finding a way of breaking supersymmetry which leaves the inflaton with a sufficiently flat potential while endowing all other moduli with steep stabilizing potentials. All standard methods of breaking supersymmetry generate, however, steep potentials, not
flat ones. One way to generate a flat inflaton potential nevertheless is to study brane-antibrane inflation in
warped backgrounds with the inflaton being identified with the brane-antibrane distance \cite{Kachru:2003sx}. Warped geometries arise in the presence of branes and fluxes. The eventual stabilization of the volume modulus, however, modifies the inflaton potential and renders it too steep for inflation unless finetuning is applied \cite{McAllister:2005mq}.

Here, we focus on an alternative mechanism to generate
inflation in M/string-theory, the multi brane inflation proposal \cite{Becker:2005sg}, \cite{ta} (see also \cite{Cline:2005ty}).
One starts with a multi inflaton scenario associating one inflaton with each brane position. The presence of several branes is indeed generically enforced by tadpole cancellation conditions. The interesting advantage of this mechanism lies in the fact that the potentials for the individual inflatons need no longer be flat. The reason is that the Hubble friction experienced by every
inflaton becomes large -- simply by increasing the number of
inflatons -- regardless of the steepness of the potentials.
This had first been pointed out in \cite{Liddle:1998jc} in the
context of 4-dimensional Friedmann-Robertson-Walker (FRW)
cosmologies based on exponential potentials which generate power-law inflation. The premise under which this mechanism operates is the suppression of strong cross-couplings among the inflatons. This suppression is given in multi brane inflation models since
interactions between non-neighboring branes which could generate cross-couplings are suppressed by longer distances. In M-theory cascade inflation \cite{ta} there is an exponential suppression of such cross-couplings since interactions between the relevant M5-branes arise from non-perturbative open M2-instantons.

In this work, after highlighting the needed ingredients of M-theory cascade inflation, we focus on the determination of its power spectrum and the resulting observable signatures. Beyond demonstrating the compatibility of the power spectrum with present cosmological constraints, we find that it exhibits {\em three distinctive signatures -- power suppression at small distances, stepwise decrease in the spectral index and oscillations in the spectrum.} The power suppression which follows in cascade inflation from M-theory dynamics might serve as an explanation for the scarceness of observed dwarf galaxies in the Milky Way halo, as suggested in \cite{Kamionkowski:1999vp}, \cite{Sigurdson:2003vy}. This is not explained by standard cosmology which overpredicts their abundance by an order of magnitude. The oscillations and stepwise decreases, on the other hand, provide a unique signature which allow to probe M-theory observationally by measuring the spectral index. It furthermore clearly distinguishes M-theory cascade inflation observationally from other string inflation models.

\section{Cascade Inflation}

Let us first provide a brief review of cascade inflation in M-theory. The starting point is M-theory on $\Orb$ \cite{Horava:1995qa}, \cite{Horava:1996ma}, also known as heterotic M-theory, compactified on a Calabi-Yau threefold $\mX$ down to an ${\cal N}=1$ supergravity in four dimensions. It is a warped compactification due to the presence of $G$-flux sourced by the 10-dimensional boundaries and additional spacetime-filling M5-branes \cite{Witten:1996mz}, \cite{Curio:2000dw}. The warped background which reduces the size of $\mX$ along the interval $\Orb$ plays an important role in cascade inflation and its exit \cite{Becker:2005sg}, \cite{ta} as well as in the possible formation of heterotic cosmic strings towards its end \cite{Becker:2005pv}. Potentials which are generated for the moduli by $G$-flux and several non-perturbative effects \cite{Curio:2001qi}, \cite{Buchbinder:2003pi}, \cite{Becker:2004gw}, \cite{Anguelova:2005jr} are too steep for inflation. It has therefore been suggested in \cite{Becker:2005sg} to use the joint effect of the interactions between several M5-branes to increase the system's Hubble friction and thus obtain inflation via the assisted inflation idea \cite{Liddle:1998jc}. Indeed, having $N>1$ M5-branes is generically required to satisfy the tadpole cancellation condition. The M5-branes are space-time filling and wrap a genus zero holomorphic two-cycle on $\mX$ to preserve the same ${\cal N}=1$ supersymmetry as the background. For simplicity, a single rigid two-cycle $\Sigma_2$ is considered. Being BPS objects, the M5-branes interact only at quantum level via open M2-instantons which stretch between them along $\Orb$ and wrap $\Sigma_2$.
For a single rigid two-cycle the sum over instantons trivializes and no integral over the moduli space of the two-cycle needs to be carried out. There are two ways to show that the distribution of M5-branes approaches dynamically an equidistant distribution along $\Orb$. The first is based on energy minimization \cite{Becker:2005sg}. Setting the K\"ahler covariant derivatives of the superpotential $D_i W=0$ with respect to the $N$ complex M5-brane position fields $Y_i$ to zero is tantamount to a partial energy minimization and implies that supersymmetry is not broken in the $Y_i$ moduli sector. Geometrically, $D_i W=0$, enforces the M5-branes to be distributed equidistantly in the large volume limit. A second way is to show that the equidistant distribution corresponds to an attractor solution \cite{ta}.

The moduli fields of interest to us are the complex M5-brane
position moduli $Y_i, \; i=1,\hdots,N$ next to the complex volume and orbifold-size moduli $S$ and $T$. Twice their real parts are denoted $y_i$, $s$ and $t$. Moreover, it's convenient to define $y = \sqrt{\sum_{i=1}^N y_i^2}$ which enters the ${\cal N}=1$ supergravity's K\"ahler potential capturing the backreaction of the M5-branes on the background geometry. For an equidistant distribution
\beqa
\text{Re}(Y_{i+1}-Y_i) = (t/2L) (x_{i+1}^{11}-x_i^{11}) \equiv (t/2L) \Delta x
\eeqa
is independent of $i$ and $\Delta x$ is identified with the inflaton. $L$ is the $\Orb$ interval size, $x_i^{11}$ the position of the $i$th M5-brane. Note that for an equidistant distribution the multi inflaton system effectively reduces to a single inflaton one. In the large volume limit -- specified by the inequality $st \gg y^2$ -- where cascade inflation occurs, the potential for the canonically normalized inflaton $\varphi \sim N^{3/2} \Delta x$ becomes
\beqa
U_N(\varphi) = U_N e^{-\sqrt{\frac{2}{p_N}}\frac{\varphi}{M_{Pl}}} \; ,
\label{Potential}
\eeqa
where
\beqa
U_N = (N-1)^2 \bigg(\frac{6M_{Pl}^4}{st^3
d}\bigg) \; , \qquad\quad p_N = N(N^2-1)
\bigg(\frac{4}{3st}\bigg) \; ,
\eeqa
are the energy scale and power parameter and $d$ the Calabi-Yau intersection number. For more details on the derivation we refer the reader to \cite{Becker:2005sg}, \cite{ta}.

The combined non-perturbative interaction of $N$ M5-branes thus leads to an exponential potential for $\varphi$ whose cosmological FRW evolution leads to power law inflation \cite{Lucchin:1984yf} for which the growth of the FRW scale factor\footnote{Note the
typographical difference between cosmic time $\mathrm{t}$ and the
modulus $t$.}
\beqa
a(\ti) = a_0 \ti^{p_N}
\eeqa
is entirely determined by the parameter $p_N$. A period of inflation occurs if $p_N>1$. It is therefore important to note that with $N$ M5-branes one finds a scaling $p_N \sim N^3$ \cite{Becker:2005sg}, \cite{ta} which easily allows $p_N$ to be larger than one and shows the generation of inflation for sufficiently many M5-branes. This can also be seen from the slow-roll parameters which decrease parameterically like $\epsilon,\eta \sim 1/N^3$ and hence become small when $N$ becomes large. In fact $N$ is bounded below and above. The bound from below follows from $p_N>1$, the prerequisite for inflation. The bound from above descends from the requirement to work in the large volume regime where $st\gg y^2$ and noticing that $y$ grows with $N$. For typical parameter values one finds $20 \le N \le 200$ \cite{Becker:2005sg}.

Let us now describe the cascade inflation phase \cite{ta}. The repulsive M2-interactions between the M5-branes cause them to spread over the $\Orb$ interval until the two outermost M5-branes hit the boundaries. The ensuing non-perturbative small instanton transition
transforms the outermost M5-branes into small instantons on the boundaries \cite{Witten:1995gx}. More precisely, the small
instantons are described by a torsion free sheaf, a singular
bundle. The singular torsion free sheaf can then be smoothed out
to a non-singular holomorphic vector bundle by moving in moduli space \cite{Ovrut:2000qi}. This process changes the topological data on the boundaries while the number $N$ of M5-branes participating in the inflationary bulk dynamics drops to $N-2$. The small instanton
transitions can be either chirality or gauge group changing \cite{Ovrut:2000qi}. We are considering the first case in which a change in the third Chern class of the visible boundary's vector bundle changes the number of fermion generations during the transition. This opens up the attractive possibility of reducing dynamically the number of generations during the cascade inflation phase, given that most compactifications exhibit a large number of generations far greater than three. Notice that for the chirality changing transition the gauge group will not change during the transition and unwanted relics are not produced.

The cascading process starts when the first two outermost M5-branes hit the boundaries and no longer participate in the bulk dynamics. The remaining $N-2$ M5-branes will continue to spread until the second most outermost M5-branes hit the boundaries in a second transition and so on. The successive stepwise drop of the number of M5-branes by two marks the cascade inflation phase \cite{ta}. Between each of these transitions we have a potential of the form (\ref{Potential}) giving power-law inflation but with stepwise decreasing values for $N$ and thus different parameters $p_N$ and $U_N$ after each transition. The cascade inflation process comes to an end when the number of M5-branes, given by
\beqa
N_m = N-2m
\eeqa
in the $m$th phase, drops below a critical value $N_K$ in the $K$th phase determined by the exit condition
\beqa
\text{exit from inflation}: \quad p_{N_K} = 1 \; .
\eeqa
Throughout the cascading process the inflaton will always be identified with the M5-brane separation and grows continuously.
The evolution during cascade inflation is thus a series of
consecutive power-law inflation phases
\begin{equation}\label{scl-fac}
a_{m}(\mathrm{t}) = a_{m}\mathrm{t}^{p_{N_m}}, \;\;\qquad
{\mathrm{t}}_{m-1}\leq \mathrm{t}\leq \mathrm{t}_m, \qquad
m=1,\hdots,K \; .
\end{equation}
Matching the scale factor at the transition times $\mathrm{t}_m$
determines the prefactors to be
\begin{equation}\label{prefac}
a_{m}=a_1
\mathrm{t}_{1}^{p_{N_1}}\left(\frac{\mathrm{t}_2}{\mathrm{t}_1}
\right)^{p_{N_2}} \left(\frac{\mathrm{t}_3}{\mathrm{t}_2}
\right)^{p_{N_3}} \ldots\;
\left(\frac{\mathrm{t}_{m-1}}{\mathrm{t}_{m-2}}
\right)^{p_{N_{m-1}}} \!\!\!
\frac{1}{{(\mathrm{t}_{m-1})}^{p_{N_m}}}
\end{equation}
The scale factor, but not the Hubble parameter, is therefore
continuous at the transition times $\ti_m$. The onset time of
inflation, $\mathrm{t}_0$, is determined by inverting the
exact power-law inflation solution for $\varphi(\ti)$ in the initial phase and noting that $\Delta x(\mathrm{t}_0)/L\ll 1$. The result is
\begin{equation}\label{t0}
\mathrm{t}_{0} \simeq \frac{2 N^2}{3M_{Pl}} \sqrt{\frac{2td}{s}}
\; .
\end{equation}
Similarly, by inverting the solution for $\varphi(\ti)$ one
obtains for the transition times \cite{ta}
\beqa\label{tr-time}
\mathrm{t}_m - \mathrm{t}_0 = \frac{1}{M_{Pl}}\sqrt{\frac{st^3
d}{6}}\bigg(\sum_{k=2}^m \frac{p_{N_k}(3p_{N_k}-1)}{N_k-1}
e^{t\big(\frac{1}{N_k-1}-\frac{1}{N_{k-1}-1}\big)} +
\frac{p_{N_1}(3p_{N_1}-1)}{N_1-1} e^{\frac{t}{N_1}}\bigg)
\eeqa
from which the number of e-foldings generated during cascade inflation follows
\begin{equation}\label{Ne}
N_e \equiv
\ln\left(\frac{a(\mathrm{t}_f)}{a(\mathrm{t}_0)}\right)
= \sum_{m=1}^{K}
p_{N_m}\ln\left(\frac{\mathrm{t}_m}{\mathrm{t}_{m-1}}\right) \; .
\end{equation}

WMAP three-year results indicate that the scalar spectral index,
$n_s$, is $0.951^{+0.015}_{-0.019}$ \cite{Spergel:2006hy}. For
power-law inflation one has $n_s = 1-2/p_N$ in the initial phase. Adopting typical values of $s=682$ and $t=14$ for the case of an unbroken hidden $E_8$ \cite{Becker:2005sg}, $N$ has to lie within the interval $61\le N\le 75$ to satisfy the spectral index constraint. Of course, the initial number of M5-branes can
be larger than this upper bound, with the proviso that the resulting $n_s$, at the scales of our Hubble radius, lies within the interval given by the WMAP data set. Taking the central value, $n_s\sim 0.951$, one finds $N=66$ M5-branes.

The scale of inflation, $M$, can be at most of order the grand unified (GUT) scale to have gravitational waves under control. Assuming instant reheating one needs about 60 e-foldings to solve
the problems of standard Big-Bang (SBB) cosmology. Mapping the
cascade inflation model, with the above values for $s$ and $t$, to GUT-scale inflation, requires $d\sim 4\times 10^5$. One might lower the required minimal number of e-foldings by either lowering the
reheating temperature, $T_{\mathrm{RH}}$, or $M$. However, for the
above choices of $s$ and $t$, lowering $M$ requires larger values for $d$ which seem to be non-generic. The details of reheating have yet to be worked out for cascade inflation, nonetheless, we assume instant reheating and therefore $T_{\mathrm{RH}}\sim M$. We should note here that the above values of $s$, $t$, $d$ and $N$ are not the only values that lead to GUT scale inflation which satisfy the theoretical and observational constraints. Surfing the landscape of parameters allows us to choose different sets of parameters. For example, one can also achieve a GUT-scale inflation by choosing $(s, t, d, N)=(6000, 11.4, 50000, 129)$ or (3000, 20, 1000, 123). As we will see, different values for these parameters determine the location of the resulting oscillations in the power spectrum. Henceforth, we will proceed with the initial values $(s, t, d, N)=(682, 14, 4\times 10^5, 66)$ although the qualitative features do not change with other choices of parameters.

Starting initially with $N=66$ M5-branes in the bulk, we find
\begin{equation}\label{t0-66}
 t_0 = 3.21 \times {10}^5
\end{equation}
in Planckian units. The total number of e-foldings is
\begin{equation}\label{Ne-66}
{N_e}_{} = 237.83,
\end{equation}
which is much larger than the number of e-foldings required to solve the horizon and flatness problems of SBB. Most of the inflationary expansion takes place within the first power-law phase in which none of the M5-branes has yet collided with the boundaries. However, we are interested in the last 60 e-foldings of expansion which are within our observable horizon.

\section{Power Spectrum of Cascade Inflation}

Inflation, besides solving the flatness and horizon problems of
standard cosmology, provides a causal mechanism to generate the
seed for large scale structures of the universe. Temperature
fluctuations of the cosmic microwave background radiation (CMBR) --
the afterglow of the BigBang -- are believed to be
generated by quantum fluctuations of the field(s)
responsible for inflation. WMAP alone indicates a flat
$\Lambda$-dominated universe with nearly scale-invariant power
spectrum with $n_s(0.05 \mathrm{Mpc^{-1}})\sim 0.95$
\cite{Spergel:2006hy}. Any viable inflationary model should be
able to produce a power spectrum compatible with these
observations.

During inflation two types of perturbations are produced: scalar
(density) perturbations and tensor perturbations (gravitational
waves). These two types of perturbations are both responsible for
the temperature anisotropy of the CMBR. Let us focus on scalar
perturbations. The evolution of Fourier components of scalar
perturbations, $u_{k}$, is known to be governed by the equation
\cite{Mukhanov:1990me}
\begin{equation}\label{scalar-per}
u''_k+(k^2-\frac{z''}{z})u_k=0 \; ,
\end{equation}
where
\begin{equation}\label{z}
z\equiv a\dot{\phi}/H \; .
\end{equation}
$u_k$ is the Fourier component of the gauge invariant Mukhanov
variable $u$, which is a combination of scalar perturbations of
the metric and inflaton \cite{Mukhanov:1990me}. $u$ is
proportional to the curvature perturbation $\mathcal{R}$ of the comoving hypersurface \cite{Lidsey:1995np}
\begin{equation}\label{u-R}
u=-z\mathcal{R} \; .
\end{equation}
In the equations above, prime (dot) denotes differentiation with
respect to the conformal (cosmic) time, $\tau$ ($\mathrm{t}$). The
solutions to the mode equation (\ref{scalar-per}) are normalized
so that they satisfy the Wronskian condition
\begin{equation}\label{Wrons}
u_k^{\ast}{u'}_k-u_k {u'_k}^{\ast}=-i.
\end{equation}
Ultimately the scalar power spectrum is defined as
\begin{equation}\label{scr-pwr}
P^{1/2}_{s}=\sqrt{\frac{k^3}{2\pi^2}}
\left|\frac{u_k}{z}\right|_{k/aH\rightarrow 0}
\end{equation}
\begin{figure}[t]\label{fig1}
\includegraphics[angle=0, scale=0.60]{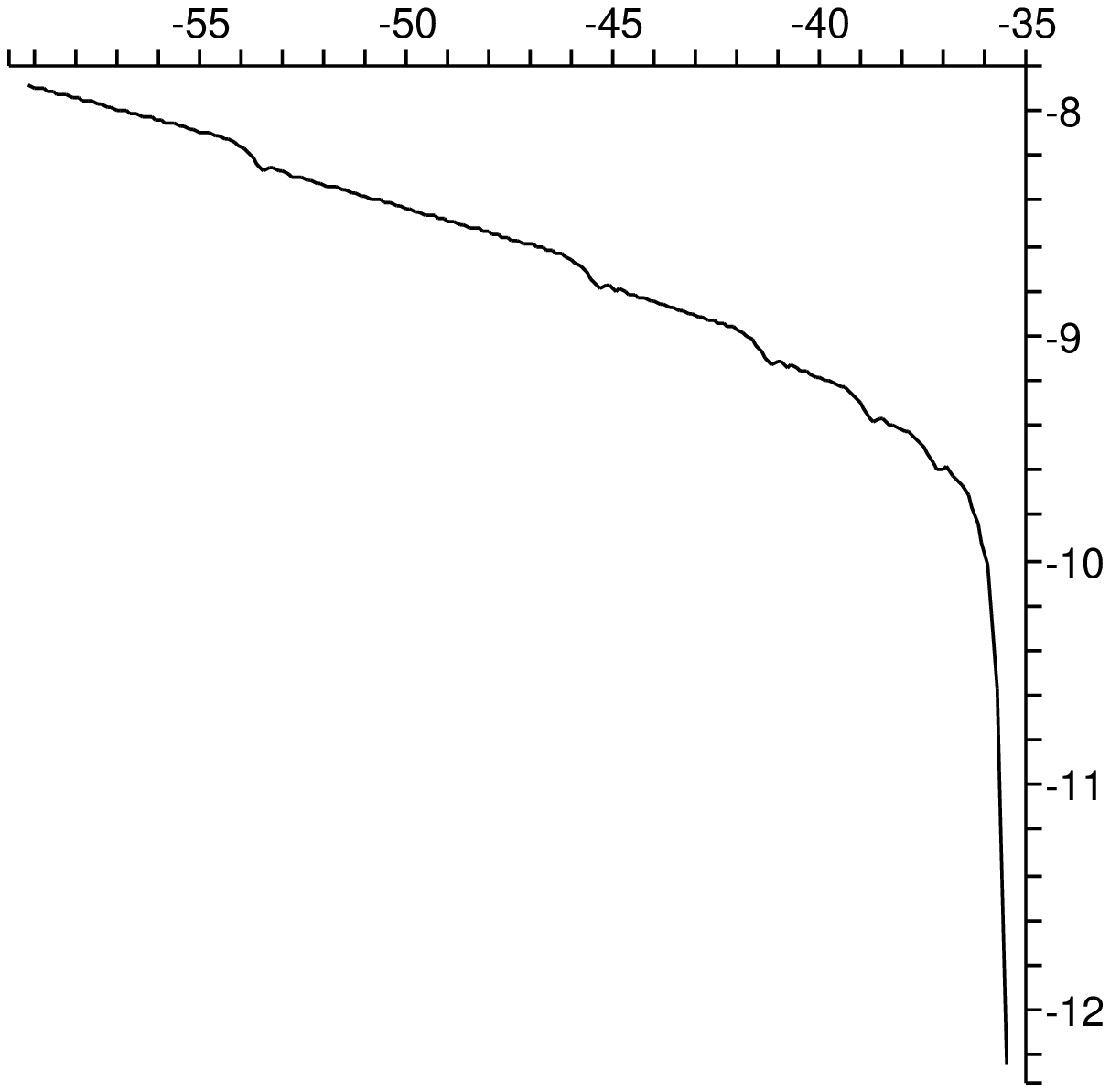}
\includegraphics[angle=0, scale=0.60]{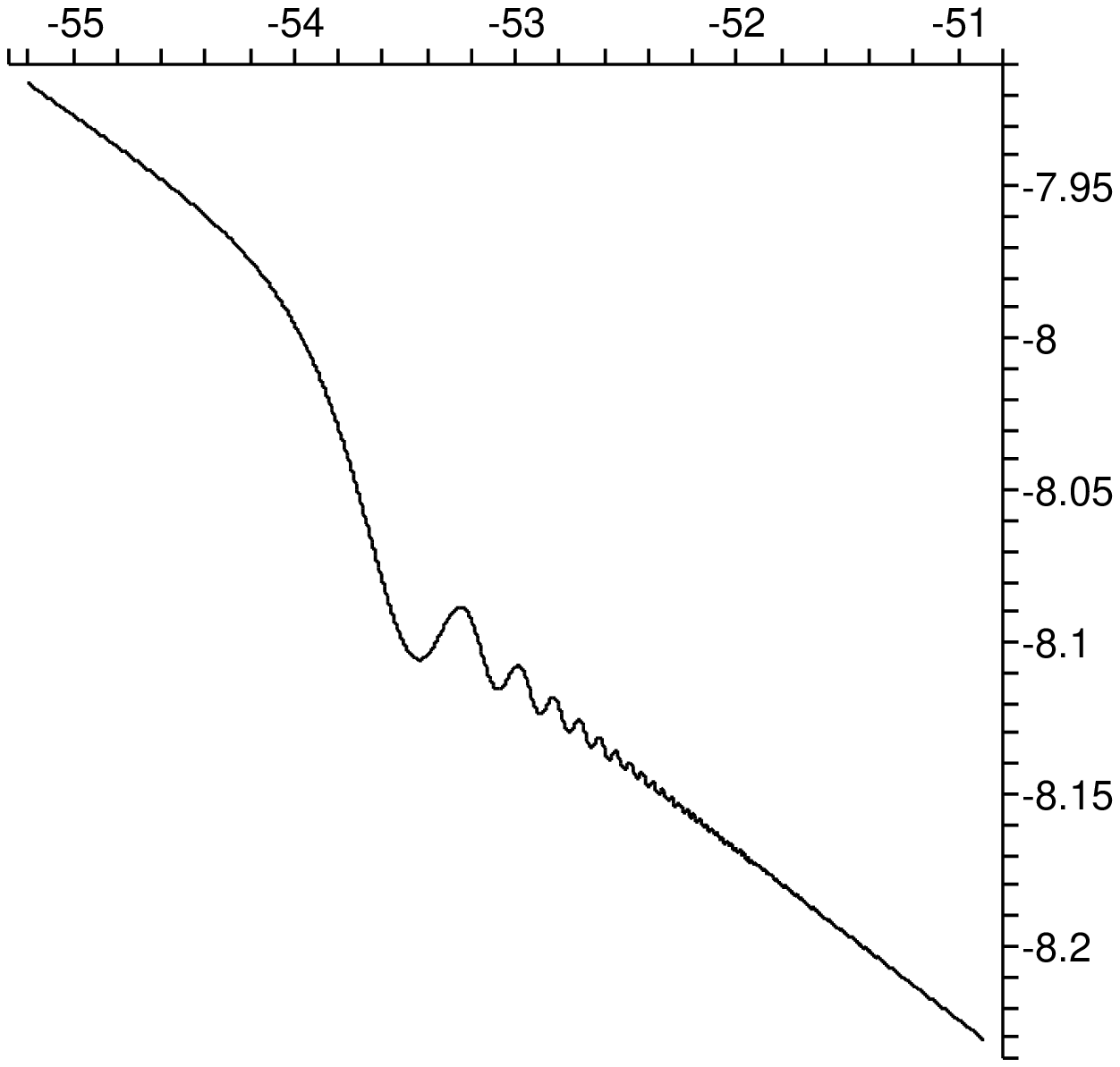}
\caption{The left graph shows the dependence of $\log P_s(k)$ on $\log
k$ for the scales that have crossed the horizon at least once
during the last $60$ e-foldings and are still outside the horizon
at the end of inflation. It clearly displays the stepwise decrease
in the amplitude of the power spectrum. The right graph shows $\log
P_s(k)$ vs. $\log k$ around the first transition. The amplitude of
oscillations decreases as $k$ increases.}
\end{figure}
The power spectrum should be evaluated in the limit where the mode
goes well outside the horizon. To recover the ordinary quantum
field theory result at very short distances much smaller than the curvature scale, we require that the mode
approaches the Bunch-Davies vacuum when $k/aH\rightarrow \infty$
\begin{equation}\label{BD-vac}
u_k(\tau)\rightarrow \frac{1}{\sqrt{2k}} e^{-ik\tau} \; .
\end{equation}

During each power-law phase equation (\ref{scalar-per})
simplifies to a Bessel equation
\begin{equation}\label{Bessel-eq}
u''_k+\Big(k^2-\frac{\nu_i^2-\frac{1}{4}}{(\tau-c_i)^2}\Big)u_k=0, \qquad
\tau_{i-1}\leq \tau \leq \tau_{i}
\end{equation}
where
\begin{equation}\label{nu}
\nu_i=\frac{3}{2}+\frac{1}{p_i-1} \; , \qquad
c_{i}=\tau_{i-1}-\frac{{{\mrm{t}}_{i-1}}^{1-p_{i}}}{a_{i}
(1-p_{i})} \; ,
\end{equation}
and $\tau_i$ is the conformal time corresponding to $\mrm{t}_i$.
The Bessel equation has the following general solution
\begin{equation}\label{pwr-mode}
u_k(\tau)=C_i(k)(c_i-\tau)^{1/2}H_{\nu_i}^{(1)}(k c_i-k
\tau)+D_i(k)(c_i-\tau)^{1/2}H_{\nu_i}^{(2)}(k c_i-k \tau)
\end{equation}
where $H_{\nu_i}^{(1)}$ and $H_{\nu_i}^{(2)}$ are the first and second Hankel functions of order $\nu_i$. Starting from
the first power-law phase and demanding that the mode satisfies
equation (\ref{BD-vac}) at the beginning of cascade inflation, one can determine $C_1(k)$ and $D_1(k)$
\begin{equation}\label{C1-D1}
C_1(k)=\frac{\pi}{2} e^{i(\nu_1+\frac{1}{2})\pi/2}\; , \qquad
D_1(k)=0 \; .
\end{equation}
\begin{figure}[t] \label{fig2}
\includegraphics[angle=0, scale=0.60]{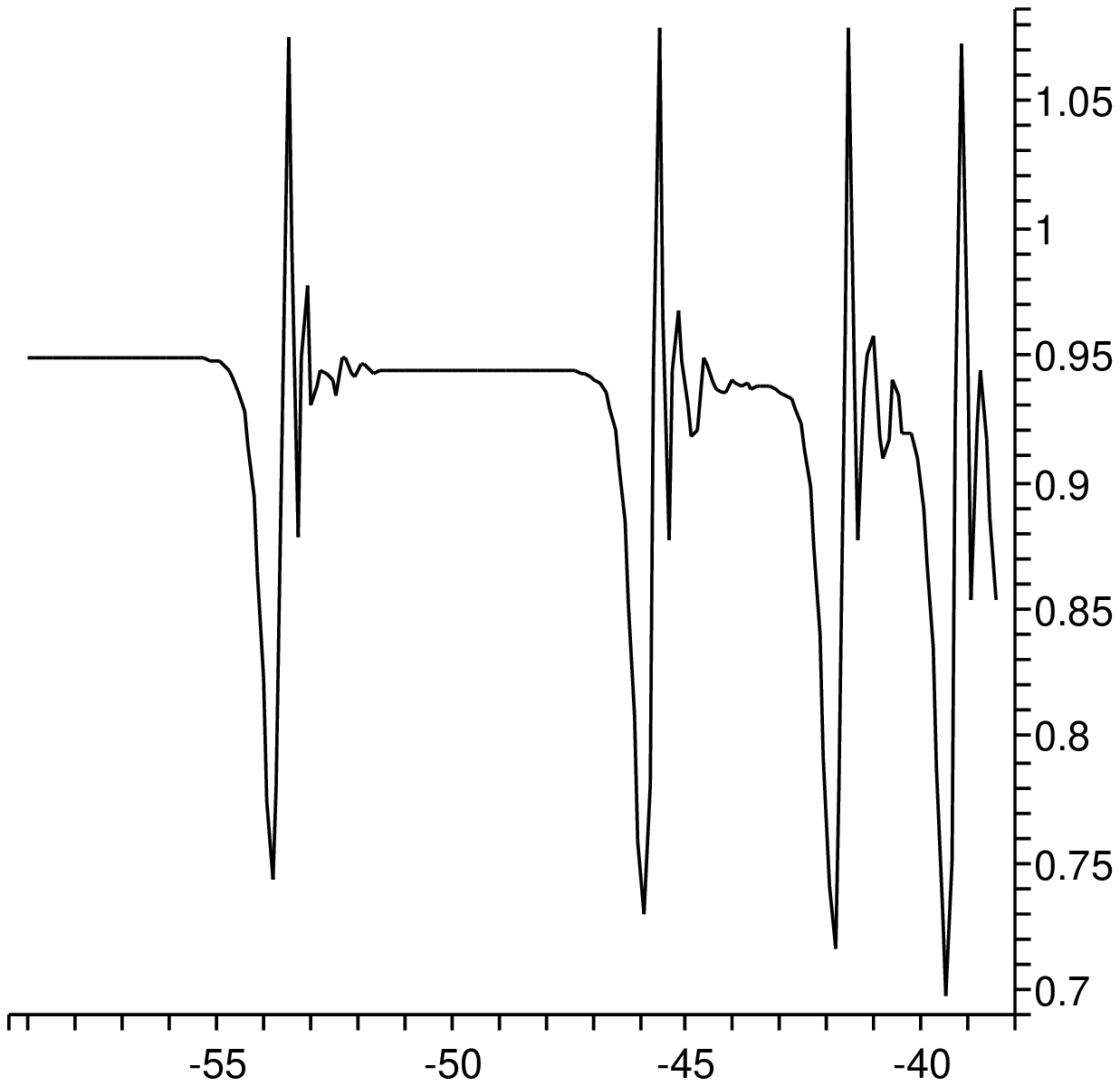}
\includegraphics[angle=0, scale=0.60]{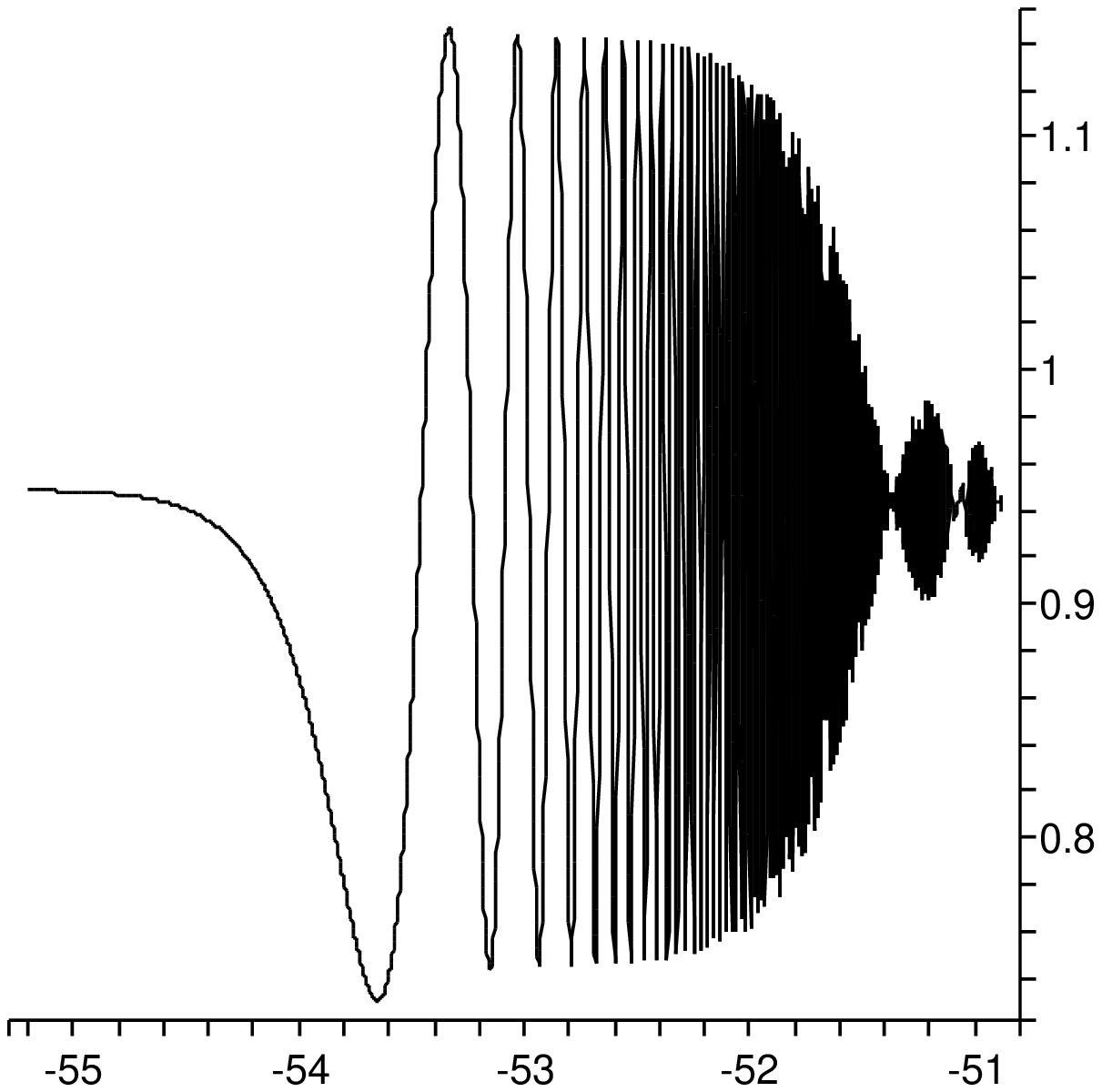}
\caption{The left graph shows the dependence of $n_s$ on $\log k$ for
the first five inflationary bouts. It clearly displays the
stepwise decrease in the spectral index. The right graph shows $n_s$
vs. $\log k$ around the first transition. The period of
oscillations decreases as $k$ increases.}
\end{figure}

Through the transition from one power-law phase to the next,
the scale factor is continuous but the Hubble parameter is not.
This happens because the potential has steps at the transitions. Potentials with step occur in supergravity motivated models of inflation where the inflaton lies within the hidden sector and is gravitationally coupled to a visible sector which contains the standard model \cite{Adams:1997de}. Spontaneous supersymmetry breaking
that occurs in the visible sector changes the mass of the inflaton
and leads to a sudden downward steps in the inflaton potential.
During inflation, the resulting change in the potential is
compensated by an increase in the inflaton kinetic energy.
Therefore the slow-roll approximations become unreliable around
the transition points \cite{Adams:2001vc}. Despite some similarity
that exists between the cascade inflation model and these supergravity motivated inflation models, there is an eminent difference. In cascade inflation, the differences in potential energy after each step are transferred to the
boundaries as the two outermost M5-branes dissolve into them via
small instanton transitions. Thus the kinetic energy of
the inflaton fields, whose role is played by the separations of the M5-branes in the bulk, will not get modified by the existence of such jumps in the potential. That is why we can still approximate the evolution by a power-law, even instantaneously after the
transitions. Since the Hubble parameter decreases whereas the scale factor remains continuous through the transitions, the size of the Hubble radius increases slightly. Therefore some modes that have just gone outside the horizon, are recaptured and start oscillating again. As we will see these oscillations will be translated to
oscillations in the power spectrum later.

Focusing on adiabatic perturbations, the three curvature
perturbations of the comoving hypersurface, $\mathcal{R}$,  is
continuous and differentiable through the transitions
\cite{Deruelle:1995kd}. This allows us to determine the $i$-th
Bogoliubov coefficients in terms of $(i-1)$-th ones and calculate
the power spectrum in the limit when all modes are far outside the
horizon. The left graph in fig.1 displays the power spectrum for the modes that have crossed the horizon at least once during the last
60 e-folds of inflation and are \tit{still} outside the horizon at
the end of inflation. Since the Hubble parameter and $p_m$, drops in each step, the total amplitude of perturbations decreases as well.
The modes that cross the horizon twice during the transitions
display oscillationary behavior. Actually, the oscillations last
for an interval of $k$ much larger than the interval crossed by the
horizon twice. As the right graph in fig.1 shows, for the first
step with approximately $7 \%$ drop in amplitude of the potential,
the oscillations last for as much as three decades of $k$. The left
graph in fig.2 presents $n_s$ vs. $\log k$ for the first five
inflationary bouts. Aside from the superimposed oscillations, the
modes pick up the value of spectral index of the bout during
which they cross the horizon. Of course, this is only true for the
first few bouts that last long enough to let the oscillations fade
away. For the last inflationary bouts which last much less than an
e-folding, this inference breaks down. The amplitudes of
perturbations suppresses significantly and we have a very red
spectrum at such scales. As the right graph in fig.2 demonstrates,
the period of oscillations in the power spectrum decreases as $k$
increases.

$N$-body simulations of structure formation use the assumption of
a scale-invariant power spectrum and predict the number of dwarf
galaxies in the Milky-way halo an order of magnitude larger than
observed. As suggested in \cite{Kamionkowski:1999vp}, a sudden
downswing of the power spectrum at small scales can explain this
discrepancy and ameliorates the disagreement between cuspy
simulated halos and smooth observed halos. Since in cascade
inflation, the power spectrum drops in each step, its value at
small scales is smaller than what a simple extrapolation of the spectrum at large scales predicts. With the above choice of parameters, the
first downturn occurs at about $0.012$ Mpc. A suppression at such a
scale ameliorates the problem of the dearth of dwarf galaxies without violating constraints from the Lyman-alpha forest. Of course, one should mention that there are also other explanations for the dearth of power at small scales, see for e.g.
\cite{Sigurdson:2003vy}.

Recently, Covi et.~al.~have tried to explain measured
deviations of WMAP three-year data from featureless power spectrum
\cite{Covi:2006ci}, using potentials with step. They found
interesting constraints on the location, magnitude and gradient of
possible step in the inflaton potential. Especially, they noticed
that current data allude to the existence of one step at the
location of WMAP low-$\ell$ glitches and one at smaller scales,
beyond the second peak. In particular, the feature at small scales
can mimic the effect of baryonic oscillations and reduce the
estimate of $\Omega_b$, baryon density. As mentioned earlier, the
freedom of the choice of number of M5-branes and other M-theory
parameters in our model, can change the location and magnitude of
the resulted oscillations. It will be interesting if one uses the
results of that paper, to derive M-theory parameters or at least
put some constraints on them.

One should also note that in generation of such oscillations from
supergravity models with hidden sector inflation, one had to
assume that symmetry breaking phase transitions happen during
inflation \cite{Adams:1997de}. In cascade inflation this
assumption will necessarily be true since the transitions are
generated by the very multi M5-brane dynamics which also drives
inflation. Hence, features in the potential will inevitably be
produced.

\section*{Acknowledgments}

We are grateful to R.~Allahverdi, R.~Brandenberger, R.~Easther, J.~Khoury, W.~Kinney, R.~Mann and R.~Woodard  for helpful
discussions. A.A.~is supported by the Natural Sciences \&
Engineering Research Council of Canada. A.K.~is supported by the
National Science Foundation under grant PHY-0354401 and the
University of Texas A\&M.

\end{document}